\documentclass[twocolumn,aps,prd,10pt,nofootinbib]{revtex4-1}
\usepackage{inputenc}
\usepackage{amsmath}
\usepackage{amssymb}
\usepackage{mathtools}
\usepackage{graphicx}
\usepackage{hyperref}
\usepackage{dsfont}
\usepackage{newtxtext}
\usepackage[cmbraces]{newtxmath}
\usepackage{slashed}
\usepackage{enumitem}

\hypersetup{
    colorlinks,
    linkcolor={blue},
    citecolor={blue},
    urlcolor={blue}
}

\allowdisplaybreaks

\graphicspath{{./}{./figs/}{../figs/}{../Figs_trial/}{../Yukawa/}}

\usepackage[dvipsnames]{xcolor}
\usepackage[normalem]{ulem}

\begin{document}

\title{%
Unconventional Gross-Neveu quantum criticality: interaction-induced SO(3)-biadjoint insulator and emergent SU(3) symmetry
}

% \begin{flushright}
%       \normalsize LTH 1253
% \end{flushright}
% \vspace{-1.2cm}

\author{Shouryya Ray}
\email{sray@cp3.sdu.dk}
\affiliation{CP3-Origins, University of Southern Denmark, Campusvej 55, DK-5230 Odense M, Denmark}

%%%%%%%%%%%%%%%%%%%%%%%%%%%%%%%%%%%%%%%%%%%%%%%%%%%%%%%%%%%%%%%%%%%%%%%
\begin{abstract}
%
%The quantum critical behavior of Dirac semimetals with SO(3) isospin has recently entered the focus of attention of the condensed matter community, because the SO(3) analogue of N\'eel order leaves behind a gapless fermionic mode, furnishing an example of what has been named ``metallic'' quantum criticality. 
I point out that Gross-Neveu theory with SO(3) isospin in three spacetime dimensions---proposed recently, for instance, as an effective description of the N\'eel transition in certain spin-orbital liquids---also hosts quantum criticality of a more exotic kind. The ordered phase breaks SO(3) spontaneously, but the SO(3)-N\'eel order parameter vanishes. The fermionic bilinear order parameter is instead a biadjoint with respect to SO(3); unlike its N\'eel cousin, it constitutes an interaction-induced insulator. Furthermore, I show that the N\'eel and biadjoint order parameters can be combined to transform as an adjoint under SU(3) symmetry; the symmetry is emergent at the critical point separating the symmetric semimetal and the biadjoint insulator, but only if the flavor number is small enough, suggesting order-parameter fluctuations and the interplay between different channels play a crucial role in stabilizing the enlarged symmetry. In candidate SO(3) spin-orbital liqiuds, thermodynamic critical exponents carry fingerprints of ``spinons''. The existence of an independent universality class in addition to the N\'eel transition opens the possibility of posing further constraints on spinon properties from thermodynamic measurements near criticality alone. %The availability of two independent universality classes against which to check theoretical predictions opens a possibility of constraining more strongly the spinons' isospin properties in candidate SO(3) spin-orbital liquids using thermodynamic critical exponents.
\end{abstract}
%%%%%%%%%%%%%%%%%%%%%%%%%%%%%%%%%%%%%%%%%%%%%%%%%%%%%%%%%%%%%%%%%%%%%%%

\date{\today}

\maketitle

%%%%%%%%%%%%%%%%%%%%%%%%%%%%%%%%%%%%
\section{Introduction}
%INTRODUCTION
%%%%%%%%%%%%%%%%%%%%%%%%%%%%%%%%%%%%
Quantum critical phenomena beyond the Ginzburg-Landau paradigm are a cornerstone of the modern theory of quantum phase transitions.\footnote{See, for instance, textbooks such as \cite{cardy96,sachdevbook,herbutbook}}
%\cite{sachdevbook,herbutbook}
Commonly, this entails degrees of freedom beyond order-parameter excitations that become soft at criticality; in a Fermi system, this role is played by fermionic degrees of freedom that ``live'' on the Fermi surface. In $d > 1$ spatial dimensions, the Fermi surface generically has co-dimension one, leading to an uncountable infinitude of such modes and an effective theory that is usually intractable analytically.\footnote{Notable exceptions where some controlled analytical progress has been possible include \cite{SSLee,Metlitski1,Metlitski2,SurLee14,SurLee15,Schlief}; see also \cite{Schlief23} for attempts to go beyond perturbation theory.} The situation improves drastically in \emph{Dirac semimetals}, where the Fermi surface consists of isolated points in reciprocal space and the fermions disperse linearly close to the Fermi points, leading to emergent Lorentz symmetry near criticality. Quantum phase transitions in such materials are described by Gross-Neveu theory in $D \coloneqq d + 1$ spacetime dimensions (GN$_D$ for short); I shall mainly focus on quasi-planar materials---of which graphene is arguably the most celebrated exponent\footnote{cf., e.g., \cite{herbut06,herbut09a,herbut09b,pujari16,ray18,singapore,ray21b,citemeR4}; an example where the ``semimetal'' is formed due to $d$-wave superconductivity in a metal is discussed in \cite{vojta00}; for a recent review of quantum criticality in Dirac semimetals, cf. \cite{boyack21}.}---so $D = 3$ unless stated otherwise. %
However, boiling down the effectiveness of GN$_3$ theory to its applicability in Dirac semimetals, or even statistical physics short-changes it somewhat. %
%However, the effectiveness of GN$_3$ theory goes not just beyond quantum criticality in Dirac semimetals, but in fact beyond statistical physics itself. %
Among other things, GN$_3$ (more generally, GN$_{2 < D < 4}$) constitutes a particularly simple, and thereby tractable, example of \emph{asymptotic safety} \cite{braun11}, which is a highly predictive candidate for the UV completion of the Standard Model coupled to Einstein-Hilbert gravity.\footnote{See \cite{Reuter:2019byg} for a pedagogical introduction and \cite{Bonanno:2020bil,Eichhorn:2022jqj} for reviews on the current state of the art.} Therein, quantum fluctuations of the spacetime metric tensor are supposed to offset the classical breaking of scale symmetry through dimensionful couplings such as the Newton coupling, in analogy with how quantum fluctuations enable the perturbatively non-renormalizable Fermi coupling to reach scale invariance in GN$_3$. Since gravity fluctuations only become strong at planckian scales inaccessible to current accelerators, quantum criticality of the GN$_3$ family is at the moment perhaps the only example of asymptotic safety with fermionic matter that can be observed in the laboratory. %
%; short of building transplanckian accelerators, quantum criticality of the GN$_3$ family is perhaps the only example of asymptotic safety with fermionic matter accessible in the laboratory.

%Although SU(2) is a somewhat natural initial point for Gross-Neveu theory with non-abelian isospin symmetry, there is no need for it to be final. This is especially so given that the quantum numbers of the fermionic quasiparticles are not restricted to be adiabatically connected to the fundamental quantum numbers of the electron. 
A significant recent application of Gross-Neveu quantum criticality is furnished by a class of novel phases of matter called spin-orbital liquids. %Nominally, these are insulating materials whose effective degrees of freedom are localized spin-orbital moments that interact via exchange interactions. However, if the interactions are frustrated (i.e., no unique energy minimum among the allowed relative orientations of moments in a unit cell), the ground state is long-range entangled with topologically non-trivial properties, and the low-energy excitations are fermionic (unlike, say, magnons in an antiferromagnet, which are bosonic); the fermionic quasiparticles (so-called spinons) can be understood as a consequence of \emph{fractionalization} of the erstwhile magnetic moments, and the spin-orbital liquid behaves 
In many regards, they behave like a Dirac semimtal, albeit one made of \emph{spinons}---fermionic quasiparticles arising due to fractionalization of the electron's spin-orbital moment---rather than elementary electrons (as would be the case in graphene) themselves. It is impossible to excite spinons in a coherent fashion: an experimentally feasible protocol like flipping a magnetic moment excites a continuum of spinons. What is comparatively feasible is to destabilize a spin-orbital liquid in favor of a conventionally ordered phase and detect the onset of said order. The exponents which describe the non-analyticity of thermodynamic observables close to criticality---called thermodynamic critical exponents---still carry ``fingerpints'' of the spinons. This makes the study of quantum criticality of the GN$_3$ family a promising component of the toolkit towards the diagnostic of novel phases of matter (in addition to all the afore-mentioned merits of GN$_{2 < D < 4}$). It has been proposed \cite{seifert20} that a potential quantum phase transition to N\'eel order in a candidate SO(3) spin-orbital liquid (e.g., in spin-orbit coupled double perovskites like Ba$_2$YMoO$_6$ \cite{natoriprl,romhanyiprl}), would be describable using GN$_3$ theory with SO(3) isospin [GN$_3$-SO(3) theory for short]; onset of N\'eel order would then correspond to spontaneous symmetry breaking (SSB) of the SO(3) isospin symmetry by the condensation of the fermion bilinear $\bar{\psi} L_a \psi$, with $L_a$ ($a = 1, 2, 3$) the generators of SO(3) in the fundamental representation. A more thorough understanding of GN$_3$-SO(3) quantum criticality is hence worth pursuing.

GN$_3$-SO(3) does have a more prominent cousin, GN$_3$-SU(2), which has been studied more extensively since it concerns graphene's putative antiferromagnetic quantum critical point \cite{janssen14,knorr18,zerf17,gracey18}. A significant difference between SO(3) and SU(2) is that SO(3) generators have a zero eigenvalue: when $\bar{\psi} L_a \psi$ condenses, the effective mass matrix of the fermions inherits this zero eigenvalue. The SO(3)-N\'eel transition is hence a semimetal-to-semimetal transition, dubbed ``metallic'' quantum criticality. By contrast, SU(2) generators have no vanishing eigenvalue, and the N\'eel-ordered phase is an interaction-induced (also called Mott) insulator.\footnote{The terminology semimetal vis-a-vis interaction-induced/Mott insulator is used somewhat liberally here. In the strict sense of those words, they refer to the presence or absence of electrically charged quasiparticles at the Fermi level. Charge here in turn refers to the physical charge of the electron, i.e., the porperty of being charged under the U(1) gauge symmetry associated with the photon. In that stricter sense, an SO(3) spin-orbital liquid is already a Mott insulator, because spinons do not carry a U(1) electric charge. For the purposes of this Article, however, any and all fermionic excitations will count towards a Fermi surface, regardless of whether these fermions have any overlap with single-particle electron states. By this logic, a state is an interaction-induced (i.e., a Mott) insulator only if it has an interaction-induced gap at the Fermi level with respect to all fermionic quasiparticles. In the remainder, the qualifiers interaction-induced and Mott are furthermore taken to be synonymous and dropped for the sake of brevity when clear from context.}

\begin{figure}[t]
 \centering
 \includegraphics[width=\linewidth]{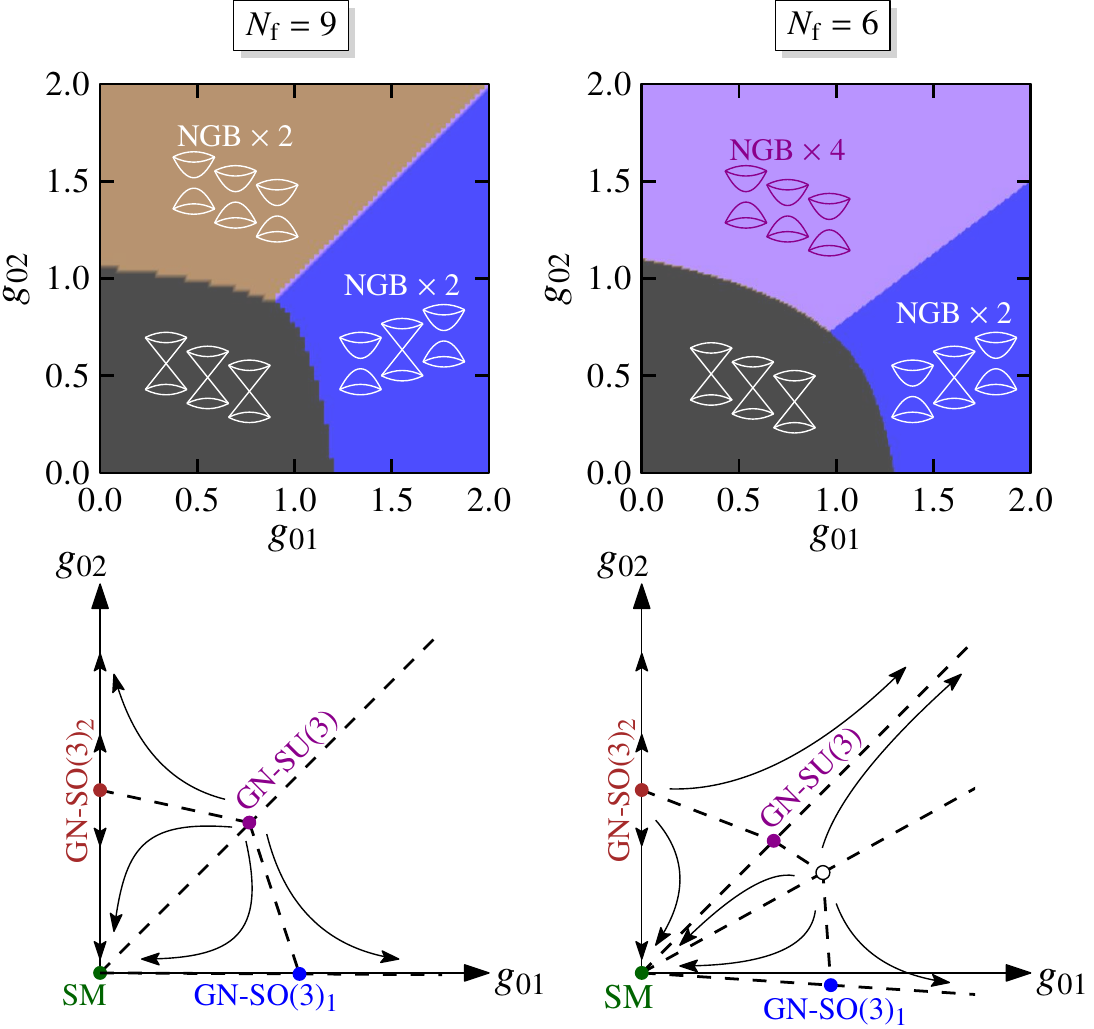}
 \caption{
 Top row: Phase diagrams as function of 4-Fermi couplings [see Eq.~\eqref{eq:Lagrangian} for definitions] for interaction-induced transitions from symmetric semimetal (SM) to Lorentz invariant SO(3) isospin-broken phases in three-dimensional Gross-Neveu theory; insets show the fermions' single-particle spectrum, along with the number of Nambu-Goldstone bosons (NGBs) where applicable. %, obtained by integrating the RG flow initialized within the $(g_{01},g_{02})$ plane. Ordered phases are characterized by divergent 4-Fermi couplings within finite flow time; the precise symmetry-broken phase is identified by comparing mutual ratios of 4-Fermi couplings at divergence with critical fixed points. %
 Bottom row: Corresponding RG phase portraits, with fixed points [projected to the $(g_{01},g_{02})$ plane] denoted by solid circles and (projections of) RG-invariant subspaces denoted by heavy dashed lines; flow lines are schematic and not to scale. Left: $N_\text{f} = 9$. In this regime, the mean-field picture is qualitatively valid. In addition to the usual N\'eel [$\coloneqq \operatorname{SO}(3)_1$] transition, there is a transition to an SO(3)-biadjoint insulator [$\coloneqq \operatorname{SO}(3)_2$] governed by the finite-$N_\text{f}$ descendant of the large-$N_\text{f}$ fixed point; the SU(3) symmetric subspace is IR-unstable. Right: $N_\text{f} = 6$. While the stability of SO(3)$_1$ remains unchanged, SU(3) symmetry emerges for $N_\text{f} < N_{\text{f},\text{cr}} \approx 6.5$ in the IR at a putative GN-SO(3)$_2$ transition. The fermions' single-particle spectrum is qualitatively the same, but robust qualitative differences in the SO(3)$_2$ vis-a-vis SU(3) scenarios include a distinct number of NGBs. There is a bicritical fixed point (white) not adiabatically connected to any mean-field limit; for $9 < N_\text{f} < 6$, it collides first with GN-SO(3)$_2$ and then GN-SU(3), realizing thus a transfer of stability between the two fixed points, before settling down between GN-SU(3) and GN-SO(3)$_1$ to form the phase boundary between the SO(3)$_1$ and SU(3) phases.
 }
\label{fig:phaseportrait}
\end{figure}

In this Article, I shall focus on a further peculiarity of GN$_3$-SO(3): unlike SU(2) [or for that matter any SU($N_\text{iso}$)], it is possible to break SO(3) isospin symmetry in Lorentz-invariant fashion at the level of fermion bilinears without giving the N\'eel bilinear $\bar{\psi} L_a \psi$---which transforms as a vector\footnote{We shall in fact see below that it is more precisely an adjoint vector, though for SO(3) the two notions are ultimately equivalent.} under SO(3)---a vacuum expectation value (vev). The resulting phase instead sees a different bilinear obtain a vev. This bilinear transforms as a \emph{bi}adjoint\footnote{Again, as discussed in more detail later, the ``biadjoint'' is equivalent to ``symmetric traceless rank-2 tensor'' for SO(3). If one likes to think of the bilinear $\bar{\psi} L_a \psi$ as a dipole moment in isospin space, then the biadjoint bilinear is, in the same vein, conducive to interpretation as a \emph{quadru}pole moment in isospin space.} under SO(3) and gaps out the fermion spectrum entirely: it is an interaction-induced insulator phase. %
My main findings can then be summarized as in Fig.~\ref{fig:phaseportrait}, in terms of phase diagrams as a function of density-density (aka 4-Fermi) couplings $g_{01}$ and $g_{02}$ associated at mean-field level with the N\'eel and biadjoint order parameters respectively [see Eq.~\eqref{eq:Lagrangian} for a more precise definition]. For large enough flavor number, an effectively single-channel description is valid. The transition to the interaction-induced SO(3)-biadjoint insulator [SO(3)$_2$ for short] is governed by a distinct critical fixed point, as is the N\'eel [$\eqqcolon \operatorname{SO}(3)_1$] transition. The SO(3)$_1$ and SO(3)$_2$ order parameters can be formally combined to form an adjoint under SU(3) for appropriate couplings, but the corresponding fixed point is then bicritical: SU(3) symmetry is unstable. This changes at small flavor number. In this regime, non-trivial interplay between fluctuations of different order parameters stabilizes SU(3) symmetry. As a consequence, the GN-SO(3)$_2$ fixed point becomes unstable, and the SO(3)$_2$ transition is governed by the GN-SU(3) fixed point instead. The interaction-induced SU(3) and SO(3)$_2$ insulators feature the same fermionic single-particle spectrum, but there are robust qualitative differences, such as in the number of Nambu--Goldstone bosons (NGBs); in addition, the SO(3)$_1$ susceptibility diverges with the same power law as the SO(3)$_2$ one at criticality due to the enlarged symmetry.

The calculations supporting these conclusions will be performed using renormalization group (RG) at one-loop. %
I provide the RG flow equations (``beta functions'') for all SO(3)-compatible 4-Fermi interactions, a Fierz-complete basis for which I have constructed. This is a technical challenge of independent interest, and already relevant to the conventional N\'eel channel\footnote{This is true for the N\'eel channel corresponding to any non-abelian isospin symmetry.} $(\bar\psi L_a \psi)^2$ studied before in \cite{seifert20,ray21}, because it is not closed under renormalization; the only non-abelian isospin GN theory for which Fierz-complete beta functions have been computed is GN$_2$-SU(2) \cite{ladovrechis23}. The present one should hence be a welcome addition to this rather sparse ``stamp collection''.% I shall then study whether the SU(3)-invariant subspace is attractive---which will turn out to (not) be the case for (large) small fermion flavor number---and elucidate differences in phenomenology arising in the two different scenarios.

%%%%%%%%%%%%%%%%%%%%%%%%%%%%%%%%%%%%%%%%%%%%%%
\section{Theory space of GN$_3$-SO(3)}
%%%%%%%%%%%%%%%%%%%%%%%%%%%%%%%%%%%%%%%%%%%%%%
To derive RG flow equations, we first need to write down a Gross-Neveu-like action which we can be sure will remain closed under renormalization. To do so amounts to writing down all 4-Fermi operators compatible with SO(3) symmetry. The Lie algebra of SO(3) is characterized by its structure constants, $[L_a , L_b] = i \varepsilon_{abc} L_c$, where $\varepsilon_{abc}$ is the Levi-Civita symbol. In the fundamental representation, the $L_a$ act on a three-dimensional vector space. Along with the identity $\mathds{1}_3$ (usually not written out in products), the SO(3) generators thus provide only 4 elements; the space of 3-dimensional matrices has dimension 9. The five missing matrices may be taken to be $Q_{ab}$, the independent components of the traceless symmetric combination 
\begin{align}
Q_{ab} = \{L_a , L_b \}/2 - (2/3)\delta_{ab}
\end{align}
where $\{\cdot,\cdot\}$ denotes the anti-commutator. The $Q_{ab}$'s commutation relation with the $L_a$ reads
\begin{align}
[L_a , Q_{bc}] = i (\varepsilon_{abd} Q_{dc} + \varepsilon_{acd} Q_{bd}).
\end{align}
Consider now an infinitesimal isospin rotation $\delta_{\epsilon_a} \psi = i \epsilon_a L_a \psi$ (and h.c. for $\bar\psi \coloneqq \psi^\dagger \gamma_0$). Under such a transformation, a fermion bilinear transforms as
\begin{align}
\delta_{\epsilon_a} (\bar{\psi}\mathcal{O}\psi) = i \epsilon_a (\bar\psi [L_a, \mathcal{O}] \psi).
\end{align}
In other words, the generators of SO(3) act on fermion bilinears in adjoint fashion; $\bar\psi L_a \psi$ transforms as an \emph{adjoint} under SO(3) and $\bar\psi Q_{ab} \psi$ as a \emph{biadjoint}. The above commutation relations further mean both $(\bar{\psi} L_a \psi)^2$ and $(\bar{\psi} Q_{ab} \psi)^2$ are invariant under SO(3) rotations. Tensoring $\mathds{1}_3$, $L_a$ and $Q_{ab}$ with Lorentz-covariant quantities leads to\footnote{Throughout, I shall work in Euclidean signature; $\slashed{\partial} \coloneqq \gamma_\mu\partial_\mu$}
\begin{align}
    \mathcal{L}_{\text{GN}_3\text{-SO(3)}} &= \bar{\psi}_i \slashed{\partial} \psi_i - \frac{g_{00}}{2N_\text{f}} (\bar{\psi}_i \psi_i)^2 - \frac{g_{10}}{2N_\text{f}} (\bar{\psi}_i \gamma_\mu \psi_i)^2 \nonumber\\
    &\hphantom{{}={}} {}- \frac{g_{01}}{8N_\text{f}} (\bar{\psi}_i L_a \psi_i)^2 - \frac{g_{11}}{8N_\text{f}} (\bar{\psi}_i \gamma_\mu L_a \psi_i)^2 \nonumber\\
    &\hphantom{{}={}} {}- \frac{g_{02}}{2N_\text{f}} (\bar{\psi}_i Q_{ab} \psi_i)^2 - \frac{g_{12}}{2N_\text{f}} (\bar{\psi}_i \gamma_\mu Q_{ab} \psi_i)^2.
    \label{eq:Lagrangian}
\end{align}
Two remarks are in order. (i) the Clifford algebra $\{\gamma_\mu , \gamma_\nu\} = 2\delta_{\mu\nu} \mathds{1}_{d_\gamma}$ is meant to be taken in its irreducible representation. In $D=3$, this means $d_\gamma = 2$. This in turn means the matrices $[\gamma_\mu , \gamma_\nu]$ which generate the spinorial part of Lorentz transformations are $\sim \varepsilon_{\mu\nu\rho} \gamma_\rho$ and do not furnish independent channels. Furthermore, the odd spacetime dimension and the irreducibility of the representation as usual combine to make $\gamma_5 = i \gamma_0 \gamma_1 \gamma_2 \sim \mathds{1}_2$. (ii) I have instated a flavor index $i = 1,\ldots, N_\text{f}/3$ with $N_\text{f} \in 3\mathds{N}_{> 0}$ in physical cases (in other words, $N_\text{f}$ counts the number of two-component fermions). It is distinct from the isospin index: the flavor structure is untouched by isospin rotations, and the theory as a whole is to be symmetric under flavor rotations. Eq.~\eqref{eq:Lagrangian} contains only the flavor-singlet combinations; flavor non-singlet 4-Fermi operators are not generated, and can in any case be re-written as a linear combination of flavor-singlet ones using Fierz identities.

%%%%%%%%%%%%%%%%%%%%%%%%%%%%%%%%%%%%%%%%%%%%%%
%\section{SU(3) symmetry.}
%%%%%%%%%%%%%%%%%%%%%%%%%%%%%%%%%%%%%%%%%%%%%%
The subspace within the GN$_3$-SO(3) theory space satisfying $g_{r1} = g_{r2}$ for $r = 0,1$ features an enhanced symmetry, viz., SU(3). This may be made manifest by gathering the $(L_a)$ and $(Q_{ab})$ together into one set as 
\begin{align}
(\Lambda_\alpha) = \left(L_a/2 , Q_A\right)
\label{eq:SU3algebra1}
\end{align}
with
\begin{align}(Q_A) = \left(Q_{23}, Q_{13}, Q_{12}, \frac12 (Q_{11} - Q_{22}), \frac{\sqrt{3}}{2} Q_{33}\right).
\label{eq:SU3algebra2}
\end{align}
The suggestive notation is to be taken seriously: the $\Lambda_{\alpha}$ obey $\operatorname{Tr}(\Lambda_\alpha \Lambda_\beta) = \delta_{\alpha\beta}/2$ and $[\Lambda_\alpha, \Lambda_\beta] = i f_{\alpha \beta \gamma} \Lambda_\gamma$, where $f_{\alpha \beta \gamma}$ are the SU(3) structure constants. (Most economically, this is seen by noticing that, up to a renumbering of $\alpha$, the $\Lambda_\alpha$ are proportional to the Gell-Mann matrices.) The $(\Lambda_\alpha)$ are hence a bona fide representation of SU(3)'s Lie algebra, in agreement with results on the $\operatorname{SU}(3) \supset \operatorname{SO}(3)$ ``missing label'' problem in representation theory.\footnote{In the mathematical literature, the inverse problem of labelling representations of SU(3) by representations of its subalgebr\ae\ is called the ``missing label'' problem, on which there is a quite extensive body of research. A classical review is \cite{Moshinsky}.}

%%%%%%%%%%%%%%%%%%%%%%%%%%%%%%%%%%%%%%%%%%%%%%%%%%%%%
\section{The interaction-induced biadjoint insulator}
%%%%%%%%%%%%%%%%%%%%%%%%%%%%%%%%%%%%%%%%%%%%%%%%%%%%%
Before proceeding to the renormalization of GN$_3$-SO(3), let us pause to consider the non-N\'eel isospin-broken phase characterized by $\langle \bar\psi L_a \psi\rangle = 0$, $\langle \bar\psi Q_A \psi\rangle \neq 0$. (I shall focus only on Lorentz-invariant phases henceforth.) In the limit $N_\text{f} \to \infty$, mean-field theory becomes exact, and the effective potential for the order parameter (OP) $\phi_A = \bar\psi Q_A \psi$ near $\phi_A = 0$ reads
\begin{align}
    V_\text{eff}(\phi_A) &= \frac{|\phi_A|^2}{2g_{02}} - \operatorname{Tr} \ln \left( \slashed{\partial} + \phi_A Q_A \right) \label{eq:tracelog}\\
    &= V_\text{eff}(0) + \frac{|\phi_A|^2}{2g_{02}} + \sum_{\sigma = 1}^3 \left(-\frac{M_\sigma^2}{2\pi^2} + \frac{|M_\sigma|^3}{6\pi}\right) + \mathcal{O}(\phi_A^4). %\frac{1}{12\pi^2}\sum_{\lambda=1}^3 \left(2 M_\lambda^2 - 2 |M_\lambda|^3 \operatorname{arccot} |M_\lambda| + 3 \ln (1 + |M_\lambda|)\right)
    \label{eq:Veff}
\end{align}
%the momentum integral appearing in the ``trace-log'' formula has been cut off at $p^2 = k_\text{UV}^2$ and all dimensionful quantities have been made dimensionless by rescaling with suitable powers of $k_\text{UV}$; an overall factor of $N_\text{f}$ in $V_\text{eff}$ has been scaled out, to wit: $V_\text{eff} \to V_\text{eff}/N$; $(M_\sigma) \coloneqq \operatorname{eigs} (\phi_A Q_A)$ are the eigenvalues of the mass matrix and particularly convenient to compute in the frame $\phi_A = (0,0,0,\phi_4,\phi_5)$, such that $\phi_A Q_A = \frac12\operatorname{diag}(\phi_4 - \phi_5/\sqrt{3} , -\phi_4 - \phi_5/\sqrt{3} , 2\phi_5/\sqrt{3})$.
Some comments regarding the evaluation of the ``trace-log'' formula in Eq.~\eqref{eq:tracelog} are in order. The operator trace $\operatorname{Tr}$ contains both an integration over all momenta as well as a trace over spinor, isospin and flavor indices. The flavor trace yields an overall factor $N_\text{f}$, which has been absorbed into the definition of the effective potential, $V_\text{eff} \to V_\text{eff}/N$. The trace over spinor and isospin indices begets a sum over all eigenvalues of the effective mass matrix, $(M_\sigma) \coloneqq \operatorname{eigs} (\phi_A Q_A)$. %
% These eigenvalues are most conveniently computed in the frame
% \begin{align}
% (\phi_A) = (0,0,0,\phi_4,\phi_5), \label{eq:phiframe}
% \end{align}
% such that
% \begin{align}
% \phi_A Q_A = \frac12\operatorname{diag}\left(\phi_4 - \frac{1}{\sqrt{3}}\phi_5 , -\phi_4 - \frac{1}{\sqrt{3}}\phi_5 , \frac{2}{\sqrt{3}}\phi_5 \right).
% \end{align}
To compute these eigenvalues, let us work in the \emph{defining} representation,\footnote{Henceforth, manifestly representation-dependent statements---such as the nature of components of a specific generator---will be made in the defining representation.} where the generators have components
\begin{align}
(L_a)_{ij} = i \epsilon_{aij}. \label{eq:defrep}
\end{align}
Then, $Q_4 \sim \operatorname{diag}(1,-1,0)$ and $Q_5 \sim \operatorname{diag}(-1,-1,2)$ are represented by real, diagonal matrices. The mass matrix $\phi_A Q_A$, on the other hand, is hermitean and traceless [being in fact an element of SU(3)'s Lie algebra, essentially due to the observation in Eqs.~\eqref{eq:SU3algebra1}--\eqref{eq:SU3algebra2} above]. Consequently, it can be diagonalized, and furthermore be written as a linear combination of the matrices representing $Q_4$ and $Q_5$. In other words, one may choose, w.l.o.g., a frame where
% \begin{align}
% (\phi_A) = (0,0,0,\phi_4,\phi_5), 
% \end{align}
% such that
\begin{align}
\phi_A Q_A = \frac12\operatorname{diag}\left(\phi_4 - \frac{1}{\sqrt{3}}\phi_5 , -\phi_4 - \frac{1}{\sqrt{3}}\phi_5 , \frac{2}{\sqrt{3}}\phi_5 \right). \label{eq:phiframe}
\end{align}
The integration over momenta is UV-divergent and needs to be cut off at some $p^2 = k_\text{UV}^2$. The $k_\text{UV}$-dependence can, however, be absorbed by measuring all dimensionful quantities in units of (suitable powers of) $k_\text{UV}$, i.e., $\phi_A \to \phi_A/k_\text{UV}$, $V_\text{eff} \to V_\text{eff}/k_\text{UV}^3$ and $g_{02} \to g_{02} k_\text{UV}$.

\begin{figure}[t]
 \centering
 \includegraphics[width=\linewidth]{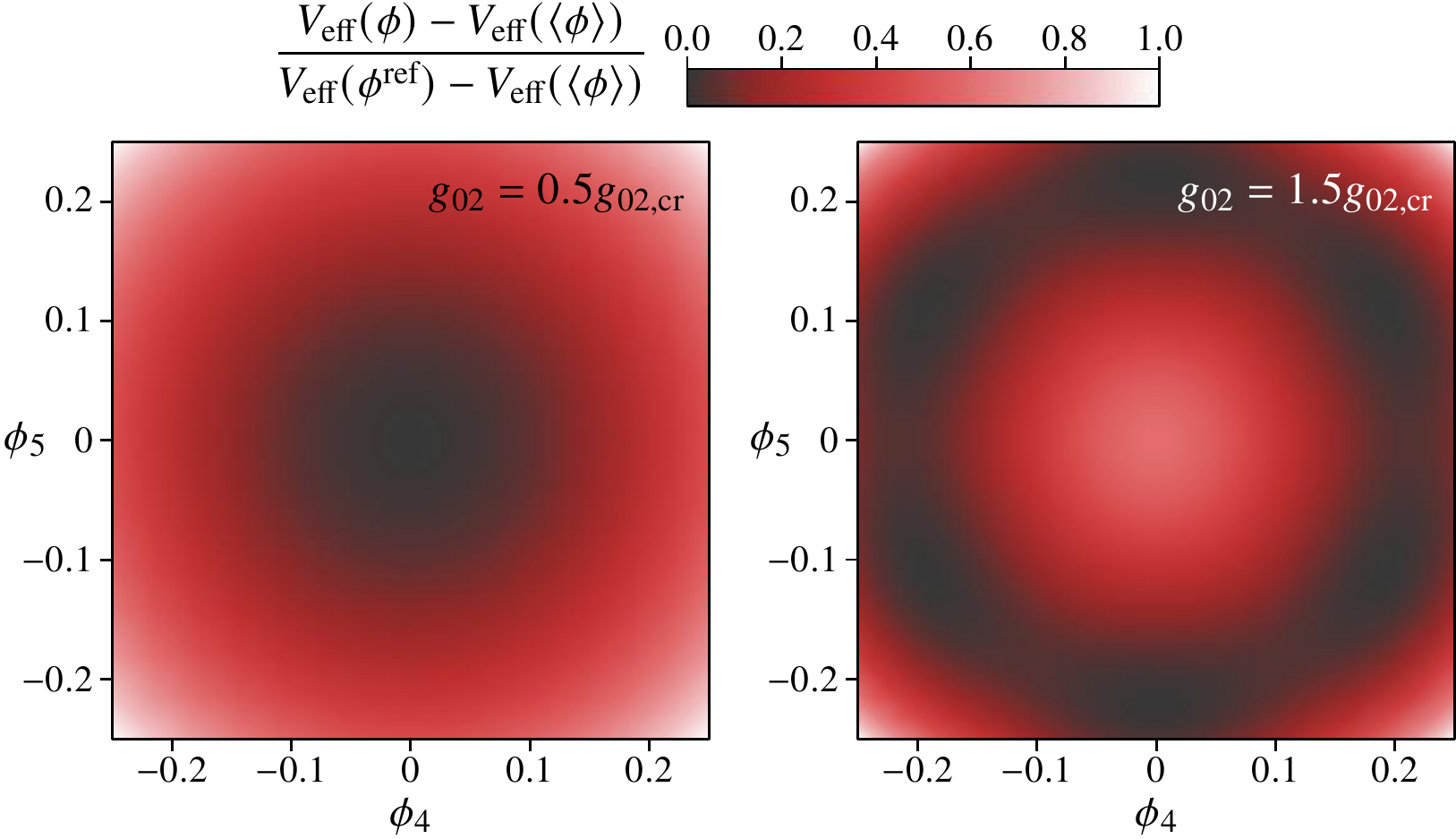}
 \caption{Plot of the effective potential of the order parameter $\phi = \phi_A Q_A$ for $g_{02} < g_{02,\text{cr}}$ (left) and $g_{02} > g_{02,\text{cr}}$ (right) in the frame $(\phi_A) = (0,0,0,\phi_4,\phi_5)$ in units of $V_\text{eff}(\phi^\text{ref}) - V_\text{eff}(\langle \phi \rangle)$, where $\phi^\text{ref} = (0,0,0,1/4,1/4)$ is an arbitrarily chosen reference point. The minimum is located at $(0,\pm v)$ up to rotations in the $(\phi_4,\phi_5)$ plane by $n\pi/6$ ($n \in \mathds{Z}$); note that the corresponding operation on $(Q_4,Q_5)$ sends the matrix $Q_5$ to one that is unitarily equivalent. The critical value of the coupling $g_{02,\text{cr}}$ can only be determined from an explicit computation, see discussion below Eq.~\eqref{eq:Veff}. All dimensionful quantities are measured in units of the UV cutoff $k_\text{UV}$, which corresponds (roughly) to the lattice constant.}
\label{fig:VeffSYMvsSSB}
\end{figure}

A plot of the effective potential in the frame defined by Eq.~\eqref{eq:phiframe} is shown in Fig.~\ref{fig:VeffSYMvsSSB} in the symmetric vis-a-vis SSB phase. One finds that $V_\text{eff}(\phi_A)$ is minimized for $\phi_4 = 0$, $|\phi_5| \propto | g - g_\text{cr} |$ if $g_{02} > g_\text{cr} = \pi^2/2$. %
The fact that the vev points in the $A=5$ direction is intuitive, since $Q_5 \equiv \Lambda_8$ is the only mass matrix that has no zero eigenvalue. Strictly speaking, the statement is true only modulo $(\phi_4 + i\phi_5) \to e^{i\frac{\pi}{3}n} (\phi_4 + i\phi_5)$ for $n \in \mathds{Z}$. The corresponding operation on $(Q_4,Q_5)$ sends $Q_5$ to one that is unitarily equivalent (i.e., has the same eigenvalues), so I shall set $n = 0$ w.l.o.g. Consequently, the fermions' single-particle spectrum is gapped out completely; from the perspective of the symmetric semimetal, this is favorable to a partial gap opening, because the reduction of density of states at the Fermi level is larger this way. Henceforth, I shall call the usual SO(3)-N\'eel phase $\langle \bar\psi L_a \psi \rangle \neq 0$ the SO(3)$_{1}$ phase; since every $L_a$ has one zero eigenvalue, this is a semimetallic phase, but with only $N_\text{f}/3$ gapless fermionic modes. The phase $\langle \bar\psi Q_A \psi \rangle \neq 0$ I shall call SO(3)$_{2}$ phase; it is an interaction-induced insulator, which I shall refer to on occasion in words as an `interaction-induced biadjoint insulator' due to the way the OP $\phi = \phi_A Q_A$ transforms under SO(3).\footnote{For SO(3), the adjoint and fundamental representations are unitarily equivalent. Consequently, there is also a unitary map between the biadjoint representation appearing here and five-dimensional representations that arise in other contexts, where they have been referred to as ``quadrupolar''/``nematic'' in the study of the spin-1 Heisenberg model (cf., e.g., \cite{mila2000,harada2002,bhattacharjee2006,laeuchli2006,tsunetsugu2006,voellwessel}), ``rank-2'' or simply ``tensor'' in the context of Luttinger semimetals (cf., e.g., \cite{IgorIgor,IgorLukas}). The present nomenclature has the advantage of making transparent the generalization from SO(3) to an arbitrary Lie group as appropriate to the context of Gross-Neveu theory with non-abelian isospin.}

%%%%%%%%%%%%%%%%%%%%%%%%%%%%%%%%%%%%%%%%%%%%%%%%%%%%%%%%%%%%%%%%%%%%%%%%
%%%%%%%%%%%%%%%%%%%%%%%%%%%%%%%%%%%%%%%%%%%%%%%%%%%%%%%%%%%%%%%%%%%%%%%

The absence or presence of a gap in the fermion spectrum is therefore one obvious way to distinguish the SO(3)$_1$ from the SO(3)$_2$ phase. This will manifest itself in thermodynamic measurements: e.g., in the SO(3)$_2$ state, the electrical conductivity will show an activated behavior as a function of temperature due to the spectral gap, whilst in the SO(3)$_1$ phase, it will follow a power law due to the leftover gapless fermionic mode.

Another way is \emph{time-reversal symmetry}. To see this, note that the internal part of the time-reversal symmetry $\mathcal{T}$ can be represented, in the appropriate basis, as complex conjugation $\mathcal{K}$: $\mathcal{T} = \mathcal{K}$. This happens to be the case in the \emph{defining} representation, where the generators have components $(L_a)_{ij} = i \epsilon_{aij}$ and are hence purely imaginary. This ensures the $L_a$, and consequently the N\'eel order parameter $\langle \bar\psi L_a \psi \rangle$, are odd under time reversal. On the other hand, the $Q_{ab}$ in the defining representation can be checked by explicit computation to be real: the SO(3)$_2$ order parameter $\langle \bar\psi Q_{ab} \psi \rangle$ is therefore \emph{even} under time reversal. This also has measurable consequences: for instance, in an infinitesimally weak external magnetic field (which is also odd under time reversal), $\langle \bar\psi L_a \psi \rangle$ will develop an infinitesimal vev; $\langle \bar\psi Q_{ab} \psi \rangle$ will not.

%%%%%%%%%%%%%%%%%%%%%%%%%%%%%%%%%%%%%%%%%%%%%%
\section{Beta functions and fixed points}
%%%%%%%%%%%%%%%%%%%%%%%%%%%%%%%%%%%%%%%%%%%%%%
The general algorithm developed by Gehring, Gies \& Janssen \cite{gehring15} ($\eqqcolon$ GGJ) provides a way to systematically derive the beta functions for a generic 4-Fermi theory at one-loop. Applying this formalism\footnote{I am grateful to K. Ladovrechis for helpful comments on the implementation of the GGJ algorithm using computer algebra.} to $\mathcal{L}_{\text{GN}_3\text{-SO(3)}}$ leads to beta functions for the dimensionless %\footnote{The dimensionless 4-Fermi couplings are defined by the rescaling $4 v_D l_1^{(\text{F})D} k^{D - 2} g_{r\ell} \mapsto g_{r \ell}$, where $k$ is the running RG scale, $v_D = [2^{D + 1} \pi^{D/2} \Gamma(D/2)]^{-1}$ comes from the angular part of the loop integration and $l_1^{(\text{F})D}$ is a dimensionless constant that encodes the (regularized) radial part of the loop integral and drops out of universal data such as critical exponents, see \cite{gehring15}. Note the wavefunction renormalization $Z_\psi = 1$ in the present approximation. Unless mentioned otherwise, I set $D=3$ when quoting final expressions such as fixed-point values of couplings and scaling dimensions. %all couplings are assumed to have been made dimensionless by rescaling with suitable powers of the running RG scale $k$.
%My sign convention is such that a \emph{negative} beta function means the dimensionless coupling will \emph{grow} towards the IR.}
4-Fermi couplings which are given by
\begin{widetext}
\begin{align}
\beta_{01} &= (D-2)g_{01} - \left(\frac{3}{D} - \frac{9}{2 D N_\text{f}}\right) g_{01}^2 + \frac{3 (12 g_{00}+36 g_{10}+12 g_{11}-15 g_{12}-5 g_{02})}{2 D N_\text{f}} g_{01} + \frac{3 (48 g_{10} g_{11}+10 g_{11} g_{12}+15 g_{12} g_{02})}{2 D N_\text{f}} \label{eq:betafcts1}\\
\beta_{02} &= (D-2)g_{02} - \left(\frac{3}{D} - \frac{3}{2 D N_\text{f}}\right) g_{02}^2 + \frac{9 (4 g_{00}+12 g_{10}+g_{12}-g_{01})}{2 D N_\text{f}} g_{02} + \frac{3 \left(48 g_{10} g_{12}+3 g_{11}^2+7 g_{12}^2+9 g_{12} g_{01}\right)}{2 D N_\text{f}} \\
\beta_{11} &= (D-2)g_{11} + \left(\frac{1}{D} + \frac{27}{4 D N_\text{f}}\right) g_{11}^2 + \frac{36 g_{00}+12 g_{10}-5 g_{12}-3 g_{01}+15 g_{02}}{2 D N_\text{f}} g_{11} + \frac{96 g_{10} g_{01}+105 g_{12}^2+20 g_{12} g_{01}+3 g_{01}^2+15 g_{02}^2}{4 D N_\text{f}} \\
\beta_{12} &= (D-2)g_{12} + \left(\frac{1}{D} + \frac{1}{2 D N_\text{f}}\right) g_{12}^2 + \frac{36 g_{00}+12 g_{10}+60 g_{11}+3 g_{01}+13 g_{02}}{2 D N_\text{f}} g_{12} + \frac{3 (16 g_{10} g_{02}+2 g_{11} g_{01}+3 g_{01} g_{02})}{2 D N_\text{f}} \\
\beta_{00} &= (D-2)g_{00} - \left(\frac{18}{D} - \frac{18}{D N_\text{f}}\right) g_{00}^2 + \frac{3 (18 g_{10}+9 g_{11}+15 g_{12}+3 g_{01}+5 g_{02})}{D N_\text{f}} g_{00} + \frac{36 g_{10}^2+3 g_{11}^2+5 g_{12}^2}{D N_\text{f}} \\
\beta_{10} &= (D-2)g_{10} + \left(\frac{6}{D} + \frac{6}{D N_\text{f}}\right) g_{10}^2 + \frac{18 g_{00}+3 g_{11}+5 g_{12}-3 g_{01}-5 g_{02}}{D N_\text{f}} g_{10} + \frac{2 (3 g_{11} g_{01}+5 g_{12} g_{02})}{3 D N_\text{f}} \label{eq:betafcts-1}
\end{align}
\end{widetext}
Here, $\beta_{r\ell} = k\partial_k g_{r\ell}$ where $k$ is the RG scale\footnote{My sign convention for the beta function is such that a \emph{negative} beta function means the dimensionless coupling will \emph{grow} towards the IR. In other words, $k \to \infty$ is the UV limit and $k \to 0$ is the IR limit.} and $g_{r\ell}$ are rescaled 4-Fermi couplings, $4 v_D l_1^{(\text{F})D} k^{D - 2} g_{r\ell} \mapsto g_{r \ell}$ in spacetime dimension $D$. The power $k^{D-2}$ simply arises from the engineering (i.e., canonical scaling) dimension of 4-Fermi operators in $D$ dimensions. The factor $v_D = [2^{D + 1} \pi^{D/2} \Gamma(D/2)]^{-1}$ comes from the angular part of the loop integration, and $l_1^{(\text{F})D}$ is a dimensionless constant that encodes the (regularized) radial part of the loop integral and drops out of universal data such as critical exponents, see \cite{gehring15}. Note the wavefunction renormalization $Z_\psi = 1$ in the present approximation. Unless mentioned otherwise, I set $D=3$ when quoting final expressions such as fixed-point values of couplings and scaling dimensions. The determination of the quadratic beta function coefficients constitutes the major technical output of this work; I have tabulated them in electronic form for download.\footnote{See Supplemental Material (SM) for electronic version of the coefficients $A_{r \ell}^{r_1 r_2 \ell \ell_1 \ell_2} = \frac12\partial^2 \beta_{r \ell} / \partial g_{r_1 \ell_1} \partial g_{r_2 \ell_2} |_{g = 0}$.}

Including degeneracies and accounting for complex solutions, there are $2^{6} - 1 = 63$ interacting fixed points $\mathcal{I}$, in addition to the Gau\ss{}ian fixed point $\mathcal{G} \colon g_{r\ell,*} = 0$. The latter corresponds to the semimetallic phase, and has no relevant directions, since the canonical dimension of 4-Fermi couplings is $[g_{r\ell}] = 2 - D = -1$. Some general facts about the interacting fixed points $\mathcal{I}$ can also be proven independently of the matrix algebra appearing in the 4-Fermi Lagrangian, as was done by GGJ; these are:
\begin{enumerate}[label=(T\arabic*)]
\item The ray $\overrightarrow{\mathcal{G}\mathcal{I}}$ is closed under RG. \label{fact:T1}
\item All interacting fixed points $\mathcal{I}$ have at least one relevant direction given by the fixed-point vector $g_{r\ell,*}$ itself; the corresponding eigenvalue of the stability matrix at $\mathcal{I}$ is $+1$. As such, this is an artefact of the one-loop approximation; however, $1/\nu = -[g_{r\ell}] = D - 2$ recovers the $D \to 2$ and $D \to 4$ limits exactly \cite{cardy96,herbutbook}. This \emph{a priori} na\"ive approximation hence tends to do unreasonably well in practice also at $D = 3$, cf., e.g., \cite{janssen14,knorr18,ray21}. \label{fact:T2}
\item If this is the unique relevant direction, $\mathcal{I}$ is called a ``critical'' fixed point. I shall notationally emphasize this by denoting such fixed points as $\mathcal{Q}$. \label{fact:T3}
\item The point $\lim_{\lambda \to \infty} \lambda\mathcal{Q} \eqqcolon (+\infty)\mathcal{Q}$ represents a stable SSB phase of matter. \label{fact:T4}
\end{enumerate}

I shall proceed with the discussion of the pertinent fixed points as follows: First, I shall restrict the beta functions to the SU(3)-invariant subspace $g_{r1} = g_{r2} \eqqcolon g_{r\text{V}}$ and consider the fixed point that has at most one relevant direction within this subspace and remains ``close'' to the $r = 0$ (i.e., Lorentz scalar) subspace. I shall then consider whether perturbations out of the SU(3) subspace [but preserving SO(3) symmetry] are relevant or not. For analytical tractability, I shall expand all quantities in powers of $1/N_\text{f}$.

%%%%%%%%%%%%%%%%%%%%%%%%%%%%%%%%%%%%%%%%%%%%%%%%%%%%%%%%%%%%%%%%%%%%%%%%%%%%%
\section{The SU(3), SO(3)$_1$ and SO(3)$_2$ fixed points and their stability}
%%%%%%%%%%%%%%%%%%%%%%%%%%%%%%%%%%%%%%%%%%%%%%%5%%%%%%%%%%%%%%%%%%%%%%%%%%%%%
The beta functions for the couplings within the SU(3) invariant subspace follow as a corollary of Eqs.~\eqref{eq:betafcts1}--\eqref{eq:betafcts-1} by setting $g_{r1} = g_{r2} \equiv g_{r\text{V}}$.
The interacting fixed point $\mathcal{Q}_\text{SU(3)}$ that describes the dynamical generation of an SU(3) breaking Lorentz scalar mass can be identified by its large-$N_\text{f}$ limit $\lim_{N_\text{f} \to \infty} g_{r\lambda,*} = \delta_{r0}\delta_{\lambda\text{V}}$. To second order in $1/N_\text{f}$, the couplings are
\begin{align}
     \mathcal{Q}_\text{SU(3)} \colon g_{r\lambda,*} &= \left(1 - \frac{1}{N_\text{f}} - \frac{8}{N_\text{f}^{2}} %+ \frac{455}{108}N_\text{f}^{-3}
     \right)\delta_{r0}\delta_{\lambda\text{V}} \nonumber\\
     &\hphantom{{}={}} {}+ \left(- \frac{3}{2N_\text{f}} + \frac{31}{4N_\text{f}^2} %- \frac{383}{216}N_\text{f}^{-3}
     \right) \delta_{r1}\delta_{\lambda\text{V}} \nonumber\\
     &\hphantom{{}={}} {}+ %\left(
     \frac{8}{3N_\text{f}^{2}} %- \frac{28}{81}N_\text{f}^{-3}
     %\right) 
     \delta_{r0}\delta_{\lambda 0} %- \frac{2}{9} N_\text{f}^{-3} \delta_{r1}\delta_{\lambda 0} 
     %\nonumber\\
     %&\hphantom{{}={}} {}
     + \mathcal{O}(1/N_\text{f}^{3}).
\end{align}
This fixed point has a unique relevant direction \emph{within} the SU(3)-invariant subspace. To study perturbations orthogonal to this space, consider $\beta_{\delta g_{r\text{V}}} = \beta_{r2} -\beta_{r1}$ with $g_{r1} = g_{r\text{V}}, g_{r2} = g_{r\text{V}} + \delta g_{r\text{V}}$. The beta functions for the dimensionless $\delta g_{r\text{V}}$ have the form $\beta_{\delta g_{r\text{V}}} = \Delta_{rr'} \delta g_{r'\text{V}} + \mathcal{O}(\delta^2)$. The eigenvalues of $\Delta_{rr'}$ at $g_{r\lambda} = g_{r\lambda,*}|_{\mathcal{Q}_\text{SU(3)}}$ are the scaling dimensions of SU(3) breaking perturbations that preserve SO(3). They are given by
\begin{align}
    \Delta_+ &= 1 + \frac{32}{3 N_\text{f}^2} + \mathcal{O}(1/N_\text{f}^{3}), \label{eq:perturbeig1}\\
    \Delta_- &= -1 + \frac{4}{3 N_\text{f}} + \frac{26}{N_\text{f}^{2}} + \mathcal{O}(1/N_\text{f}^{3}). \label{eq:perturbeig2}
\end{align}
The eigenvalue $\Delta_+$ corresponds to a perturbation primarily towards $r = 1$ (= Lorentz vector) channels. On the other hand, $\Delta_-$ corresponds to a perturbation predominantly within the Lorentz scalar subspace. For $N_\text{f} \to \infty$, it is \emph{negative}: SU(3) symmetry is \emph{not} emergent at criticality.

This result is intuitive enough to understand, once embedded within the SO(3) theory. In the strict large-$N_\text{f}$ limit, all channels decouple. As a corollary of \ref{fact:T2}, the number of relevant directions of an interacting fixed point in this limit is then equal to the number of non-vanishing couplings. The fixed point $\mathcal{Q}_\text{SU(3)}$ is, in SO(3)-terms, the fixed point $g_{r\ell , *} = \delta_{r0}(\delta_{\ell 1} + \delta_{\ell 2}) + \mathcal{O}(1/N_\text{f})$; it is hence actually bicritical. %
In the regime of large but finite $N_\text{f}$, the two critical fixed points are instead
\begin{align}
    \mathcal{Q}_{\text{SO(3)}_1} \colon g_{r\ell,*} &= \left(1 + \frac{3}{2 N_\text{f}} + \frac{3}{4 N_\text{f}^2}\right)\delta_{r0}\delta_{\ell 1} \nonumber\\
    &\hphantom{{}={}} {}- \left(\frac{1}{4 N_\text{f}} + \frac{43}{48 N_\text{f}^2}\right)\delta_{r 1}\delta_{\ell 1} \nonumber\\
    &\hphantom{{}={}} {}+ \frac{1}{6 N_\text{f}^2} \delta_{r1}\delta_{\ell 0} + \frac{1}{4 N_\text{f}^2}\delta_{r1}\delta_{\ell 2} \nonumber\\
    &\hphantom{{}={}} {}+ \mathcal{O}(1/N_\text{f}^3), \\
    \mathcal{Q}_{\text{SO(3)}_2} \colon g_{r\ell,*} &= \left(1 + \frac{1}{2 N_\text{f}} + \frac{1}{4 N_\text{f}^2}\right) \delta_{r0} \delta_{\ell 2} \nonumber\\
    &\hphantom{{}={}} {}- \left(\frac{5}{4 N_\text{f}} - \frac{65}{48 N_\text{f}^2} \right)\delta_{r1} \delta_{\ell 2} + \mathcal{O}(1/N_\text{f}^3).
\end{align}
Their physics is characterized by their $N_\text{f} \to \infty$ behavior,
\begin{align}
\mathcal{Q}_{\text{SO(3)}_\ell} \colon \lim_{N_\text{f} \to \infty} g_{r \ell' , *}|_{Q_{\text{SO(3)}_\ell}} = \delta_{r0} \delta_{\ell \ell'};
\end{align}
they are (the finite-$N_\text{f}$ descendants of) the critical fixed points describing the interaction-induced transition to the N\'eel phase ($\ell = 1$) and the biadjoint insulator ($\ell = 2$) respectively. The phase diagram\footnote{To compute the phase diagram as a function of $g_{01}$ and $g_{02}$, the RG flow is initialized within the $(g_{01},g_{02})$ plane and integrated. Ordered phases correspond to a divergence of 4-Fermi coupling(s) within finite flow time $t_\text{SSB}$. The precise nature of the ordered phase is characterized using \ref{fact:T4}, i.e., by comparing the (finite) ratios of 4-Fermi couplings at $t_\text{SSB}$ with those obtained at the critical fixed points.} that emerges corresponds to the one shown in the left panel of Fig.~\ref{fig:phaseportrait}.

However, once $N_\text{f}$ is decreased, the different channels begin to interact non-trivially with each other, and the eigenvalues for explicit SU(3)-breaking perturbations obtain corrections. Whether $\Delta_-$ will be driven towards irrelevance, and whether $\Delta_+$ will remain positive, cannot be answered \emph{a priori}, but can only be decided by an explicit computation, the result of which is Eqs.~\eqref{eq:perturbeig1}-\eqref{eq:perturbeig2}. Thus, we arrive at the main result of this paper: for $N_\text{f} \lesssim N_{\text{f},\text{cr}}$, $\mathcal{Q}_\text{SU(3)}$ becomes \emph{stable} with respect to SO(3)-invariant breaking of SU(3) symmetry. In the above approximation, $N_{\text{f},\text{cr}} \approx 2 + \sqrt{30} \approx 7.5$.\footnote{Working to all orders in $N_\text{f}$ by solving the fixed-point equations and determining the stability matrix eigenvalues numerically only gives modest corrections, $N_{\text{f},\text{cr}} \approx 6.5$.}

It is now worth asking, what erstwhile critical SU(3) non-invariant fixed point(s) the flow is attracted from. The two natural candidates are precisely the $\mathcal{Q}_{\text{SO(3)}_1}$ and $\mathcal{Q}_{\text{SO(3)}_2}$ discussed above. It turns out, that $\mathcal{Q}_{\text{SO(3)}_1}$ obtains no further relevant directions. Even at small $N_\text{f}$, it continues to govern the universal behavior of the N\'eel transition. In fact, numerically solving the beta functions shows that the fixed-point couplings $g_{r\ell,*}|_{Q_{\text{SO(3)}_1}}$ for $r \neq 0$ or $\ell \neq 1$ remain at least an order of magnitude smaller than $g_{r\ell,*}|_{Q_{\text{SO(3)}_1}}$. This makes it plausible that the leading thermodynamic critical exponents derived in \cite{ray21} using a battery of higher-order field theory methods but in an effectively one-channel setting for the GN$_3$-SO(3)$_1$ transition should at most receive corrections at the 10\,\% level once subleading channels are included. Computations beyond 10\,\% accuracy, however, should ideally account for the subleading channels if they are to be reliable and internally consistent. On the other hand, $\mathcal{Q}_{\text{SO(3)}_2}$ develops a second relevant direction. It ceases to describe the interaction-induced transition from the symmetric semimetal to the biadjoint insulator phase. The flow towards $(+\infty)\mathcal{Q}_{\text{SO(3)}_2}$ is instead re-directed towards $(+\infty)\mathcal{Q}_{\text{SU(3)}}$, leading to the phase diagram on the right panel of Fig.~\ref{fig:phaseportrait}.

\begin{figure*}
\includegraphics[width=\textwidth]{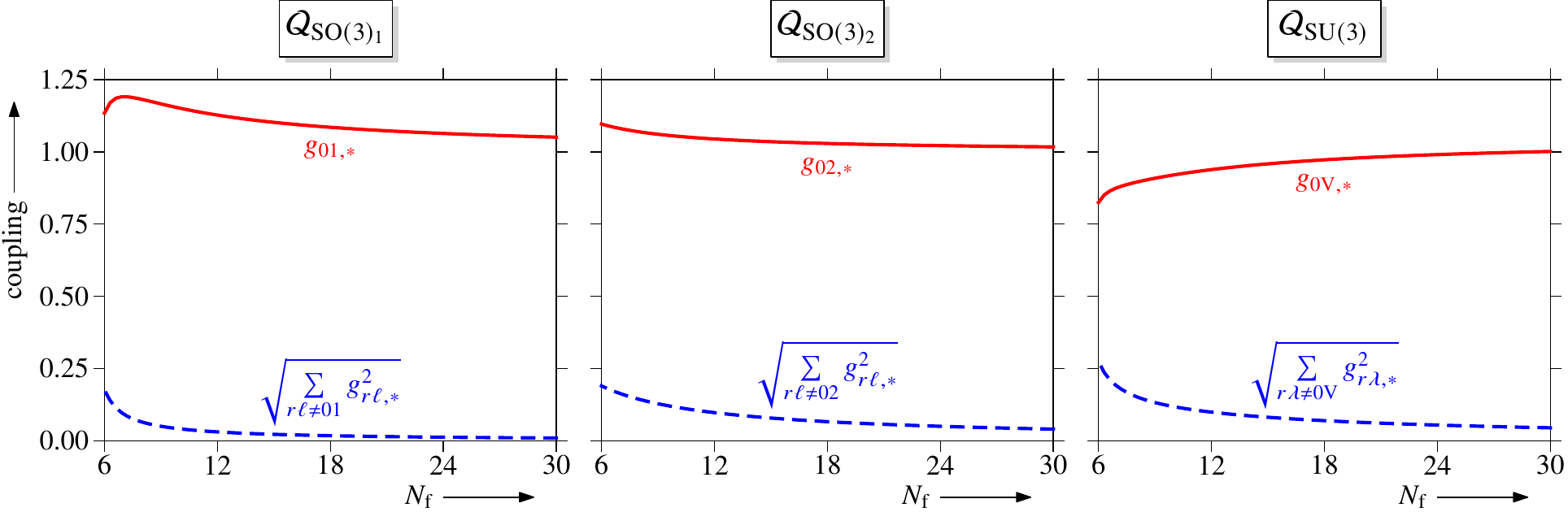}
\caption{Values of 4-Fermi couplings at fixed points $\mathcal{Q}_{\text{SO(3)}_\ell}$ ($\ell = 1,2$) and $\mathcal{Q}_{\text{SU(3)}}$, showing the nearly single-channel nature of these fixed points persists down to small $N_\text{f}$.} \label{fig:FPcoupls}
\end{figure*}

There are some technical curiosities concerning this exchange of fixed-point stability between $\mathcal{Q}_{\text{SO(3)}_2}$ and $\mathcal{Q}_{\text{SU(3)}}$, which I wish to mention by way of closing this section: $\mathcal{Q}_{\text{SO(3)}_2}$ and $\mathcal{Q}_{\text{SU(3)}}$ live in different RG-closed subspaces of the full GN$_3$-SO(3) theory space. As a result, they cannot collide with each other\footnote{I thank L. Janssen for alerting me to this.}. There is instead a bicritical fixed point $\mathcal{B}$ not adiabatically connected to the mean-field limit, which first collides with the critical $\mathcal{Q}_{\text{SO(3)}_2}$ at an $N_{\text{f}} = N_{\text{f},\text{cr}}' > N_{\text{f},\text{cr}}$ and exchanges stability with it, before proceeding to do the same with $\mathcal{Q}_{\text{SU(3)}}$ at $N_{\text{f}} = N_{\text{f},\text{cr}}$. When the dust settles, (i) $\mathcal{B}$ remains bicritical and comes to lie in the sector spanned by the rays $\overrightarrow{\mathcal{G}\mathcal{Q}_{\text{SU(3)}}}$ and $\overrightarrow{\mathcal{G}\mathcal{Q}_{\text{SO(3)}_1}}$ [i.e., it separates the SO(3)$_1$ phase from the SU(3) insulator], and (ii) $\mathcal{Q}_{\text{SO(3)}_2}$ and $\mathcal{Q}_{\text{SU(3)}}$ have effectively exchanged stability despite never entering each other's RG-closed subspace. There is hence also an intriguing theoretical scenario for $N_{\text{f},\text{cr}} < N_\text{f} < N_{\text{f},\text{cr}}'$, where both $\mathcal{Q}_{\text{SO(3)}_2}$ and $\mathcal{Q}_{\text{SU(3)}}$ are bicritical. Alas, it turns out that $3 \lfloor N_{\text{f},\text{cr}}/3 \rfloor  < N_{\text{f},\text{cr}} < N_{\text{f},\text{cr}}' < 3 \lceil N_{\text{f},\text{cr}}/3 \rceil$: though interesting in its own right, this regime is not realizable for any physical flavor number $N_\text{f} \in 3\mathds{N}_{> 0}$. %\footnote{By the same token, one might ask where the flow towards $(+\infty)\mathcal{Q}_\text{SU(3)}$ comes from for $g_{02}/g_{01} < 1$. It turns out a bicritical fixed point emerges from the complex plane to act as the missing source. The elucidation of its characteristics (along with its erstwhile complex conjugate) is beyond the scope of the present discussion.}. The picture that emerges in the small-$N_\text{f}$ regime is shown in the right panel of Fig.~\ref{fig:phaseportrait}. I shall now proceed to discuss the phenomenological consequences of this change of RG flow topology.
As an aside, I should point out here that the identification of the fixed points $\mathcal{Q}_{\text{SO(3)}_\ell}$ and $\mathcal{Q}_{\text{SU(3)}}$ relies primarily on the fact that they are connected adiabatically (in fact, differentiably as a function of $N_\text{f}$) to the large-$N_\text{f}$ limit, where the associated ordered phase can be identified trivially due to the single-channel nature of the fixed points. I have checked numerically, that down to the lowest $N_\text{f}$'s considered here, the fixed points remain approximately single-channel, in the sense that there is a clearly dominant channel inherited from the strict large-$N_\text{f}$ limit; the remaining channels continue to be subleading in the sense that the fixed-couplings in those channels continue to be much smaller. This is shown in Fig.~\ref{fig:FPcoupls}. Though an unambiguous identification of the ordering tendencies associated with these fixed points would require an analysis of the susceptibility exponents (essentially, the anomalous dimensions of all fermion bilinears with up to two SO(3)-covariant isospin indices), the relative sizes of the fixed point couplings make it reasonable to expect that the ordering preferred at mean-field level (i.e., for $N_\text{f} = \infty$) continues to be valid at small $N_\text{f}$. Finally, the fixed point $\mathcal{B}$ which collides with $\mathcal{Q}_{\text{SO(3)}_2}$ at $N_{\text{f}} = N_{\text{f},\text{cr}}'$ and $\mathcal{Q}_{\text{SU(3)}}$ at $N_{\text{f}} = N_{\text{f},\text{cr}}$ moves significantly in theory space as a function of $N_\text{f}$. Consequently, it is impossible to read off the leading ordering tendency by an inspection of the fixed-point couplings themselves, but would require one to perform a separate systematic study of the renormalization of the fermion bilinears themselves. This is left for future work.

%%%%%%%%%%%%%%%%%%%%%%%%%%%%%%%%%%%%%%%%%%%%%%%%%%%%%%%
\section{The symmetry-broken insulating phase: SO(3)$_2$ vs SU(3)}
%%%%%%%%%%%%%%%%%%%%%%%%%%%%%%%%%%%%%%%%%%%%%%%%%%%%%%%
The single-particle fermion spectrum in the SSB phase is not qualitatively different in the SU(3) vis-a-vis SO(3)$_2$ insulator phases. In both cases, the mass matrix is $\sim Q_5 \equiv \Lambda_8$, having no zero eigenvalue and thus leaving no gapless fermions post-SSB. However, there are sharp differences, already at the \emph{qualitative} level, that manifest themselves in other observables, of which I shall discuss two examples.

\begin{enumerate}[label=(\roman*)]

\item \emph{Number of Nambu-Goldstone bosons (NGBs).} If the symmetry is only SO(3), we need only consider the three generators $L_a$. Among them, only $L_1$ and $L_2$ do not commute with $Q_5$ ($\equiv \Lambda_8$). Consequently, the biadjoint insulator $(+\infty) \mathcal{Q}_{\text{SO(3)}_2}$ has two NGBs, just like the SO(3)$_1$ phase. On the other hand, at $(+\infty) \mathcal{Q}_\text{SU(3)}$, the emergence of SU(3) symmetry means every $\Lambda_\alpha$ that fails to commute with $\Lambda_8$ ($\equiv Q_5$) supplies an NGB. Direct computation reveals there are hence \emph{four} NGBs in the SU(3) insulator.

\item \emph{Divergent susceptibilities.} The large-$N_\text{f}$ universality classes are characterized by one of the SO(3)$_\ell$ susceptibilities diverging near criticality, while the other one remains finite. In the SU(3) transition, however, both the SO(3)$_1$ and SO(3)$_2$ order parameters transform as components of an adjoint vector under SU(3). Consequently, in addition to a divergent SO(3)$_2$ susceptibility, one will now have a divergent SO(3)$_1$ susceptibility at the transition to the interaction-induced insulator with spontaneously broken isospin symmetry.
\end{enumerate}

Both examples constitute robust signatures that may be expected to survive beyond the present approximation. %and on the other be readily visible in appropriate experiments.
Furthermore, (ii) is practically interesting: a completely gapped fermion spectrum in the SSB phase accompanied by a divergent SO(3)$_1$ susceptibility is a ``smoking-gun'' signature one can search for in experiments to verify whether SU(3) isospin symmetry emerges or not at the transition to the insulating SSB phase.

%%%%%%%%%%%%%%%%%%%%%%%%%%%%%%%%%%%%%%%%%%%%%%
\section{Outlook: Some Applications}
%%%%%%%%%%%%%%%%%%%%%%%%%%%%%%%%%%%%%%%%%%%%%%
A recent Letter \cite{zihong} presented Quantum Monte Carlo simulations of an $N_\text{f} = 12$ GN$_3$-SO(3)$_1$ transition using a $t$-$V$ model with Hamiltonian of the form
\begin{align}
    H &= H_t + H_V \\
    H_t &= t \sum_{\langle i j \rangle} c^\dagger_{i\sigma} c_{j\sigma} + \text{H.c.} \\
    H_V &= V \sum_{i} \left(c^\dagger_{i\sigma} L_a^{\sigma \sigma'} c_{i\sigma'}\right)^2
\end{align}
where $c^\dagger_{i\sigma}$ ($c^\dagger_{i\sigma}$) creates (annihilates) a fermion with SO(3) quantum number $\sigma$ at site $i$ of a honeycomb lattice, and $\langle i j \rangle$ denotes nearest-neighbor bonds. (The cited reference also had a bilayer index, which I shall neglect here.) Interestingly, a further quantum phase transition from the N\'eel semimetal phase to an interaction-induced SSB insulator phase was found. The latter phase was diagnosed to have vanishing N\'eel order parameter and a spontaneously broken U(1) symmetry. Assuming the absence of a narrow coexistence phase or a weak first-order transition, this would be a classically forbidden order-to-order quantum phase transition, and has been suggested to be a candidate for a so-called deconfined quantum critical point. The findings I have presented raise the intriguing possibility of tuning an SO(3) Dirac system through a further interaction-induced N\'eel-to-SSB insulator transition by %using the bicritical SU(3) fixed point instead.
including the perturbation
\begin{align}
    H_{V'} &= 4V' \sum_{i} \left(c^\dagger_{i\sigma} Q_A^{\sigma \sigma'} c_{i\sigma'}\right)^2,
    \label{eq:HVprime}
\end{align}
which has the same symmetries as $H_V$, and tuning close to the SU(3)-invariant point. This continuous transition would be governed by a conventional bicritical endpoint. Critical exponents measured at this transition would be an interesting point of comparison: if they are indeed very different from those one could measure at the putative deconfined quantum critical point of \cite{zihong}, then this would point towards a genuinely different underlying mechanism. On the other hand, if the two sets of exponents are close to each other, one may be motivated to look for whether other aspects of the observed phenomenology may in fact be describable by a conventional mechanism.

Furthermore, using $H_t + H_{V'}$ as a starting point (i.e., $V/t \ll V'/t$), one should be able to also simulate a pure GN$_3$-SO(3)$_2$ transition (note that $N_\text{f} = 12$ is well above the critical flavor number below which SU(3) becomes emergent at the insulating transition). Furthermore, $N_\text{f} = 12$ is large enough for the $1/N_\text{f}$ to provide reliable theoretical benchmarks already at low orders. Interestingly, for the GN$_3$-SO(3)$_1$ transition, (some of) the critical exponents determined numerically in \cite{zihong} showed significant deviations from the theoretical prediction. It would be interesting to see whether a similar discrepancy presents itself also for the GN$_3$-SO(3)$_2$ transition.

Beyond numerical experiments, there are also candidate materials that have been suggested to realize SO(3) quantum spin-orbital liquid (= spinon semimetal) ground states \cite{natoriprl,romhanyiprl,seifert20}. Critical exponents governing thermodynamic observables of quantum phase transitions from a spin-orbital liquid phase to magnetically ordered phases provide insight into the nature of the otherwise experimentally elusive spinons, such as their isospin symmetry. The present findings suggest one could do so in at least one ``orthogonal'' direction---viz., the biadjoint insulator---in addition to the N\'eel phase. If the two interaction-induced transitions can be tuned independently of each other, there would be twice as many independent quantities [i.e., $\nu_{\text{SO(3)}_1}, \gamma_{\text{SO(3)}_1}, \nu_{\text{SO(3)}_2}, \gamma_{\text{SO(3)}_2}$] which may be measured in principle. Such combined measurements would allow one to constrain the admissible isospin content of spinons in putative quantum spin-orbital liquids more sharply than when considering the N\'eel transition in isolation. This is a tantalising prospect and suggests a more concerted attack in future investigations from several angles:
\begin{enumerate}
    \item On the theoretical side, it would be worthwhile to subject the critical exponents of the GN$_3$-SO(3)$_2$ and GN$_3$-SU(3) universality class to a battery of more sophisticated techniques, along the lines of what was done for GN$_3$-SU(2) in \cite{knorr18,zerf17} or GN$_3$-SO(3)$_1$ in \cite{ray21}.
    \item On the phenomenological side, one would like to identify what microscopic interaction gives birth to the $g_{02}$ 4-Fermi coupling. (Note that here, one needs to identify an interaction between spin-orbital moments, unlike the setting of \cite{zihong} where one can start with itinerant fermions and Eq.~\eqref{eq:HVprime} yields the desired construction.) In spin-1 magnets, spin-nematic phases are stabilized by a biquadratic generalization of the Heisenberg interaction \cite{harada2002,laeuchli2006,tsunetsugu2006,voellwessel}. For the spin-orbital model of \cite{seifert20}, the explicit form of the corresponding generalization, though just as symmetry-allowed, is less clear, at least to me.
    \item On the ab initio side, once the pertinent microscopic spin-orbital interaction has been identified, it would be interesting to compute what is the microscopic value of this new interaction in candidate SO(3) spin-orbital liquids such as Ba$_2$YMoO$_6$ and twisted-bilayer $\alpha$-RuCl$_3$.
\end{enumerate}

\begin{acknowledgments}
I thank K. Ladovrechis, T. Meng, G. P. de Brito, M. M. Scherer, L. Janssen and A. Eichhorn for discussions and collaboration on related projects. I am moreover indebted to the last three of the aforementioned for carefully reading and providing helpful comments on earlier versions of this manuscript. Further enlightening discussions with %
A. A. Christensen and G. J. Jensen %
are gratefully acknowledged. This work has been supported by a research grant (29405) from VILLUM FONDEN and partially by the Deutsche Forschungsgemeinschaft (DFG) through the Walter Benjamin program (RA3854/1-1, Project id No. 518075237).
\end{acknowledgments}

%%%%%%%%%%%%%%%%%%%%%%%%%%%%%%%%%%%%%%%%%%%%%%%%%%%%%%%%%%%%%%%%%%%%%%%
% BIBLIOGRAPHY: FOR USE WITH BIBTEX
%%%%%%%%%%%%%%%%%%%%%%%%%%%%%%%%%%%%%%%%%%%%%%%%%%%%%%%%%%%%%%%%%%%%%%%
\bibliographystyle{longapsrev4-2}
\bibliography{GN-SO3-multiloop2}

\end{document}